\documentclass{article}
\usepackage{setspace}
\usepackage{graphicx}
\usepackage[super]{natbib}
\usepackage{amssymb, latexsym}
\usepackage{amsmath}
\usepackage{amsthm}
\usepackage{bm}
\usepackage[mathscr]{eucal}
\usepackage{enumerate}
\usepackage{multirow} 
\usepackage[center]{caption}
\usepackage{endnotes}

\usepackage{geometry}

\usepackage{authblk}
\usepackage{abstract}

\title{Sound impact of simple viscoelastic damping changes due to aging and the role of the double bentside on soundboard tension in a 1755 Dulcken harpsichord}

\author{Rolf Bader}
\author{Niko Plath}
\author{Patrick Kontopidis}

\affil{
	Institute of Systematic Musicology\\ 
	University of Hamburg\\ 
	Neue Rabenstr. 13, 20354 Hamburg, Germany, R\_Bader@t-online.de}

\begin{document}

		\maketitle

\begin{abstract}
The sound perception of wood aging is investigated on a Dulcken harpsichord of 1755 from the Museum of Applied Arts in Hamburg, Germany using a Finite-Difference Time Domain (FDTD) model of the harpsichords soundboard. The soundboard thickness was measured on the instrument at 497 positions during strings being deattached and used in the model. Impulse responses were taken on the instrument to estimate the present internal damping by calculating the T60 decay time and used as a model input. By varying the internal damping from this measured damping as a logarithmic decrement, impulse responses were simulated at 52 string positions on both, the 8' and 4' bridge. To estimate the changed sound brightness due to changed internal damping, spectral centroids were calculated from the simulated impulse responses. A dependency of brightness change due to aging on string position was found, where the lower strings have higher brightness, as expected, while the higher strings have decreased brightness. This counterintuitive finding is caused by the frequency-dependent filter effect of changed damping. Future studies need to incorporate viscoelasticity to differentiate this effect further. Furthermore, the attachment of the 8' string to the outer instead of the inner wall, a characteristic feature of Dulcken harpsichords, is investigated using a 3D Finite-Element Method (FEM) model simulation of the whole instrument. No considerable changes on the soundboard tension were found compared to an attachment of the 8' strings to the inner wall, pointing to another reason for this special construction. 
\end{abstract}

\section{Introduction}

Acoustically, the harpsichord has similarities and differences compared to the piano, the guitar, or the harp.

Like the piano it has at least one bridge through which the vibrations of the strings are transferred to the soundboard. The modern piano has this bridge in the center of the soundboard which was a major constructive change introduced by Theodore Steinweg in the mid-19$^{th}$ century following his patent\cite{Plath2017}\cite{Plath2019}. The bridge or bridges of the harpsichord are placed more near the edge of the soundboard, in the case of the Dulcken mainly the 8' bridge. The resulting acoustic difference is a spatially blurred sound with the piano while the harpsichord sound has a much more clear localization of the played string in terms of the area the sound is emitted from the soundboard\cite{Beurmann2010}. This is clearly perceived when standing at the side of piano and harpsichord. While a clearer localization helps listeners to follow a polyphonic texture typical for Renaissance or Baroque music, the Romantic sound ideal of the 19$^{the}$ century is supported by the placement of the bridge in the center of the soundboard with a spatially blurred timbre.

Like the modern piano, the soundboard of the harpsichord typically changes its thickness from about 5 mm with the bass strings to a bit more than 2 mm in the treble region, although considerable thickness differences appear. The cutoff bar at the lower left side of the soundboard, again common with pianos and harpsichords makes the area behind the cross-bar only slightly vibrating. This leaves the main vibrating area to the bridge side with both instruments. But while the piano has ribs all over the soundboard, the harpsichords normally only has ribs behind the cross-bar. Still the bridge side of the harpsichord is additionally strengthened by the 4' support or hitchpin rail, a thickened area of the soundboard on which the 4' strings are attached and which need to counteract the string tensions. Additionally, with the piano, the grain direction of the wood is in a 45 degree angle to the keyboard while the ribs are glued in 90 degrees with respect to this grain direction to support the soundboard statically, and compensate the imbalance of orthotropic material properties. The grain direction of the harpsichord is along the long end, in a 90 degree angle to the keyboard, although the 1745 Dulcken (US.W.si 315758), 1747 Dulcken (B.A.mv 1967.001.115), and the 1755 Dulcken (B.B.mim 1608) have grain angles of about 80 degrees.

This piano setup of cross-bar, bridge, and ribs, together with the gluing of the soundboard into the boundary support leads to a fundamental difference of the sound energy distribution from the string into the soundboard compared to the harpsichord. Strings distribute their energy to the bridge in small portions over one time period of the fundamental string frequency\cite{Bader2013}\cite{Bader2021}\cite{Bader2005}\cite{Beurmann2010}. During the rest of this periodicity, the string is more or less inactive with respect to sound transmission. This knocking spreads a wave from the bridge point all over the soundboard. With the piano, this spreading is nearly perfectly circular, where energy distributes into all directions away from the impact point with equal speed\cite{Plath2017}. With the harpsichord, the spreading of the impulse on the bridge is at first traveling along the bridge at high speed and only slowly driving the areas next to the bridge. This traveling along the bridge is caused by the bridge being much more thick compared to the soundboard. As thick structures have a much higher speed of sound compared to thin structures, the energy is much faster traveling along the bridge than on the soundboard. This behaviour would therefore also be expected with the piano, as also there the vibration is much faster along the bridge. Still the special distribution of the parts on the piano soundboard and its overall shape lead to this very round distribution which seems to be the main reason for the more blurred spatial radiation of pianos compared to harpsichords.

This wave traveling behaviour of the harpsichord necessarily makes the impact point of the string onto the bridge the point of maximum energy. Therefore, we can expect the soundboard to have most energy around the string impact point. This makes the harpsichord similar to the harp where the point the string is attached to the soundboard is also the point of maximum vibrational energy\cite{Waltham2008}. Such a favourable design leads to an increased loudness of instruments as when the soundboard impedance\endnote{Impedance is the resistance a vibrating body shows when driven by another body. It is calculated as Z(f) = v(f) / F(f), so the vibration velocity v a body has when driven by an external force F. This resistance is frequency dependent, so each frequency has its own impedance. When the impedance of a soundboard driven by a string is high, only little energy is transferred from the string to the soundboard. Therefore, the sound is low in volume but lasts long. Vice versa, if the impedance is low, much energy is transmitted in the same amount of time, so the soundboard vibrates stronger but the string energy is sucked by the soundboard faster and so the sound has not too much sustain.} is lowest at these points more energy can be transferred from the string into the soundboard per time interval and therefore making the instrument louder.

The harpsichord also shares a common feature with the guitar and the harp, namely the soundhole not present with pianos. For a soundhole to make sense, a closed air volume needs to be present, resulting in a Helmholtz resonance. This Helmholtz resonance acts like a mass-spring system, where the air inside the volume acts like a spring as it is compressed by the air around the soundhole acting like a mass\cite{Fletcher2000}. As the speed of sound around 343 m/s at room temperature is much slower than the speed of structure borne sound, the resonance frequency of a Helmholtz resonator is very low, with guitars around 100 Hz\cite{Bader2005}. This allows low string frequencies to be resonated strongly by such instruments which would not have a bass frequency region too prominent when only the wooden soundboard would resonate. Modern pianos are so big that their soundboards are capable to resonate even at very low frequencies, 50 Hz with the piano, while harpsichord soundboards are not. To counterbalance this, harpsichords have a closed box below the soundboard with a soundhole, just like guitars. Such air volumes with an opening have a Helmholtz resonance which is low in frequency and therefore contribute bass to the sound.

The present paper addresses the aspect of instrument aging. It is known that aged wood changes its material parameters\cite{Obataya2017}. Noguchi et al. report an increase of sound velocity and a decrease of internal damping while the rigidity ratio between Young's and shear modulus\endnote{Young's modulus is a material parameter of hardness. It is defined as the proportional constant between a stress applied to a body and the strain the body shows due to this stress. Larger Young's modulus leads to higher wave velocity on the soundboard and therefore to higher eigenfrequencies. Shear modulus is related to the Young's modulus but here the stress applied to the body is causing a shearing, not a contraction like with the Young's modulus.} was found unchanged\cite{Noguchi2008}. When seasoning wood for six months, again an increase of sound speed and a decrease of internal damping was found, which disappears after exposure to humid conditionsr\cite{Obataya2020}. The exposure of wood to UV light leads to a considerable increase of sound speed and a decrease of density\cite{Gurau2023}. Differences in density relations between grain and cross-grain directions were found with Cremona violins considerably different from modern violins, still this seems to be caused by the use of wood grown under harsh climate conditions leading to a such a relation\cite{Stoel2008}. The reason for this variations is still under debate. Increased crystalization of cellulose microfibrils was suggested but not found empirically\cite{Noguchi2008}\cite{Yokoyama2009}. An effect clearly present is the decomposition of hemicellulose leading to increased stability and reduced hygroscopicity\endnote{Hygroscopicity is the ability of a body to absorb moisture, so water, into its structure. If this ability is increased, moisture present in the air is more easily diffuses into the wood and therefore alters the sound of the instrument stronger.}. This effect is reversible when exposing wood to humid conditions\cite{Obataya2017}.

The strings used historically are now well investigated. Underwood\cite{Underwood1992} found a brass and and iron string in an Italian historical harpsichord of 1612 where the brass string is expected to be part of the original stringing and the iron string part of a reconstruction of the instrument at the end of the 18$^{th}$ century. Using x-ray spectroscopy the brass string was found to have only 15\% zinc where modern brass strings have about 20-30\% zinc content. A lesser zinc content is expected to lead to a slightly reduced linear density and therefore to slightly less tension of the strings onto the soundboard. Brass strings with low zinc content are also expected to be less bright in sound. The present Dulcken instrument is stringed with brass strings in the bass and iron strings in the middle and upper key range and expected to be modern alloys.

The speed of sound in a wooden plate not only depends on the Young's modulus and density of a plate but also on its thickness. Therefore, instrument builders can adjust variations of these parameters by altering the plate thickness. Alternations of internal damping are more complex to achieve. Therefore, the effect of decreased damping on the sound of the Dulcken is investigated more closely.

A straightforward assumption is that the sound power radiated from the instrument is getting stronger when internal damping is decreased. Indeed, measurements of plates with varying damping found such a positive correlation\cite{Ono1996}. These investigations only considered the lowest eigenmode of the plate while driving the plate right at this eigenmode. Still, harpsichords soundboards, like those of stringed instruments in general, are driven by strings with forced oscillations often not at the eigenfrequencies of the plates but somewhere between these eigenmodes. When damping is low, string frequencies not matching the soundboard frequencies are not resonated in the soundboard and therefore not radiated loudly. In other words, the soundboard acts as a filter which only supports string frequencies matching its eigenfrequencies. Still, when damping is increased, string frequencies driving the soundboard around its eigenfrequencies, not perfectly matching them will still be radiated to a considerable amount. These frequency regions around the soundboard eigenfrequencies are the measured Q (quality factor) value of the resonance with which normally the damping of this resonance is measured with.

As a consequence it was found that plates with increased internal damping radiate louder than plates with lower damping in theory\cite{Loredo2011}\cite{Kou2015} and with the piano soundboard\cite{Bader2022}. Therefore, we would expect a decrease of damping with aged wood of the Dulcken harpsichord to lead to a sound radiation which is less bright compared to modern instruments. Using a physical model of the instrument the internal damping can be easily altered and the effect quantified. We are aware that internal damping is highly complex and subject to frequency-dependent viscoelastic damping\cite{Holz1973}. Therefore in this paper only the overall effect is investigated while a deeper understanding of internal damping is subject to future studies.   

The harpsichord is investigated only poorly in terms of measurements and physical modeling. Beurmann et al. found a clear cut in the eigenmodes along the bridge using microphone array measurements\cite{Beurmann2010}. Using the same technique also performing a Finite-Element Method (FEM) calculation of a historic harpsichord, additionally the internal stress due to stringing is discussed\cite{LeMoyne2012}. Eigenmode calculations have also been performed discussing the influence on FEM parameters on eigenfrequencies\cite{Larisch2020}. Physical modeling of the harpsichord only considering the sound and not the instrument geometry have also been performed\cite{Välimäki2004}\cite{Penttinen2006}. To our best knowledge, no study discusses aging or the impact of a double bentside as is subject to the present paper.

\section{Method}

The Dulcken harpsichord build in Antwerp 1755 in the collection of the Museum of Applied Arts in Hamburg, Germany (D.Hkm 2000.526) is investigated using physical modeling of its geometry. The harpsichord geometry was taken from several sources, a 3D scan of the external geometry, an X-ray picture giving insight in its interior, a CAD model build from the X-ray picture, measurements of the soundboard thickness, visual insight into the instrument with a gyroscope inserted at the left open front side of the instrument below the soundboard, measurements of simple plate geometry and thicknesses at the instrument, as well as a drawing of a similar Dulcken instrument dated 1745 in the Kunsthistorisches Museum in Vienna  (A.W.km 726). Especially the CAD derived from the X-ray and the 3D scan gave precise positions of the 8' and the 4' bridge, the attachment points of the strings at their endpoints, as well as placement of the cross-bar and support bars around the soundhole.

\subsection{Building the Dulcken model}

Combining all the methods was necessary, as each of them only gives a certain amount of data. So e.g. to our best knowledge, the thickness distribution of the soundboard, performed by the authors of this paper in cooperation with the MK\&G during string replacements is the first thickness distribution measurement of a historical harpsichord ever made with such precision. It was performed using a MAG-ic device which measures the plate thickness using the magnetic hall effect. This allows for a precision of 1/100$^{th}$ of a millimeter. Although the thickness precision of the instrument is high, the spatial precision of the measurement was restricted to visual inspection. The device was placed at 497 positions on the instrument where each position was recorded by visual inspection on the soundboard. As visual cues, measurements in mm from boundaries or edges were used, as well as typical points like positioning the device next to each string point. Therefore, we estimate that the spatial precision is of $\pm$ 1 mm uncertainty. Furthermore, the repair of the instrument in 1990 by Beurmann, known from the inscription 'Fenster geöffnet am 10. Nov. 1990 Eigentum: Dr. Beurmann, Hasselburg' [Window opened at 10. Nov. 1990 Ower: Dr. Beurmann, Hasselburg] at the 4' hitchpin rail shows extensive tapering of the soundboard, especially in the center and treble regions. The thickness measurements included the tapering at these positions and is therefore not the original thickness. Other deviations from a smooth distribution are not always known but expected to come from glue, metal nails disturbing the magnetic field, or the like. 

\begin{figure}
	\centering
	\includegraphics[width=1\linewidth]{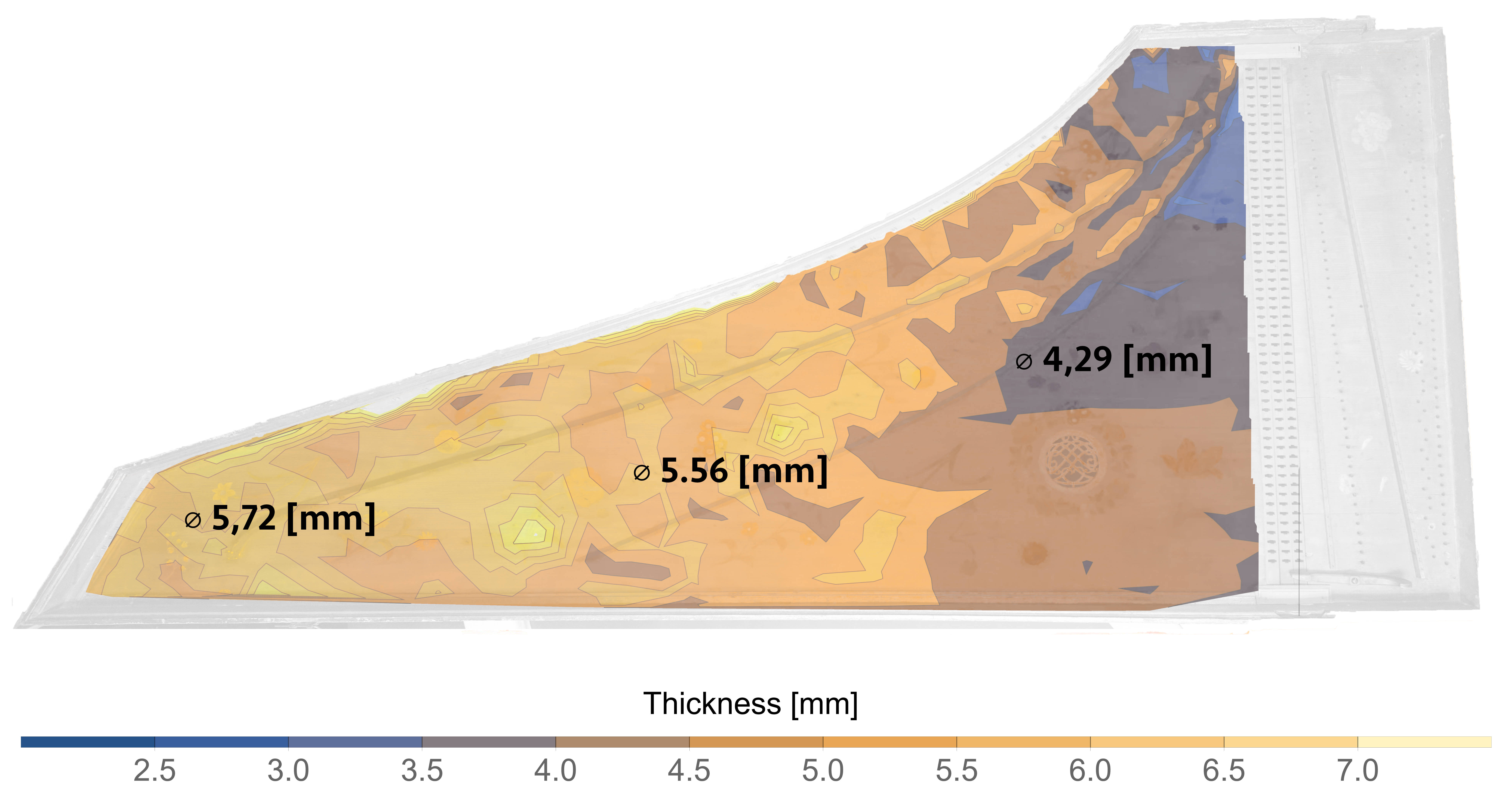}
	\caption{Measured thickness distribution of the Dulcken harpsichord of 1755 from the Museum of Applied Arts in Hamburg measured using a magnetic hall effect thickness measurement device. In the treble region the soundboard is about 2.4 mm, in the bass region about 5.7 mm with continuous thinning along this way. The contour plot consists of 497 measurement points, the measurement was performed while strings were deattached. At the region of the cross bar around the soundhole the thickness is more or less constant around 4.3 mm. Larger local variations of thickness are caused by the tapering of the soundboard during a repair in 1990, or additional glue.}
	\label{fig:cembalocontourplotfinalv5}
\end{figure}

Taking all this into consideration, the main finding remains clear: the soundboard thickness in the treble region is around 2.4 mm and increases up to 5-6 mm in the bass region, while the soundboard part behind the cutoff bar has a constant thickness of about 4.3 mm. This is shown in Fig. \ref{fig:cembalocontourplotfinalv5} with minimum thickness around 2.4 mm and maximum values at 7.5 mm. Typical values are plotted in the figure at three different positions. Modern harpsichord builders most often follow the general rule to have the lowest thickness at the treble and the highest at the bass side of the soundboard. The measurements at the Dulcken of 1755 now confirmed that this pattern was already applied by Dulcken and probably by other instruments makers of this time. 

From these geometrical details a Finite-Difference Time Domain (FDTD) was generated suitable for simulating soundboard vibrations including internal damping and being driven at different positions.

\subsection{Finite-Difference Time Domain (FDTD) Model}

The FDTD model assumes a discretized geometry to consist of a regular grid of nodal points for which the differential equations of motion are solved (for mathematical details of such models please refer to\cite{Bader2005}\cite{Bader2022}). To arrive at a modeling spectrum up to 20 kHz, a minimum grid size is needed which again depends on material and geometrical parameters. In the case of the Dulcken a spatial discretization of 1 cm was found to allow for frequencies up to about 30 kHz to be represented, allowing overhead in terms of geometry or material changes performed. This choice enforces a temporal discretization for the calculation to converge, in the present study a temporal step size of $\Delta t = 5 \times 96 000$ s was used, corresponding to five times the typical audio sampling rate of 96 kHz.

Only the soundboard was modeled in the FDTD study as a mathematical plate model to allow for estimating varying damping due to aging. Additionally, attached to the soundboard the bridges, cross-bar and the four ribs were modeled as geometries on their own as bars. This allows the use of different directions of Young's modulus present with the bridges and ribs compared to the soundboard. It also allows for estimating the filter function of these elements. Furthermore, as the cross bar and the ribs have a thickness of about 2 cm and the bridges have a thickness of 1.9 - 1.2 cm, strictly speaking the plate theory does not hold for such geometries as plate theory only holds for thin plates. The soundhole was omitted in the FDTD study as for now the enclosed air volume below the soundboard was omitted, too, and estimated analytically. The thickness of the 4' hitchpin rail was included as a thickness deviation of the soundboard. This is reasonable due to the plate theory used in the physical model\cite{Bader2005} is only valid for thin plates which still holds when including the hitchpin rail but is assumed to no longer be valid when adding the bridges. The soundboard and the bridges and ribs are coupled as a force-force interaction in both directions.

Since the precision of the FDTD grid of 1 cm in x- and y-direction is way enough to arrive at 30 kHz modeling frequency, the pins locations of the strings onto the 8' bridge was taken the same for the two 8' strings. The 4' only has one string per key. The instrument has 52 keys, resulting in 2 $\times$ 52 = 104 string points for both, the 8' and the 4' bridges.

Although the model also contains strings, for the investigation performed they were omitted and only impulse responses at the string points on the bridges were modeled.

\begin{figure}
	\centering
	\includegraphics[width=1\linewidth]{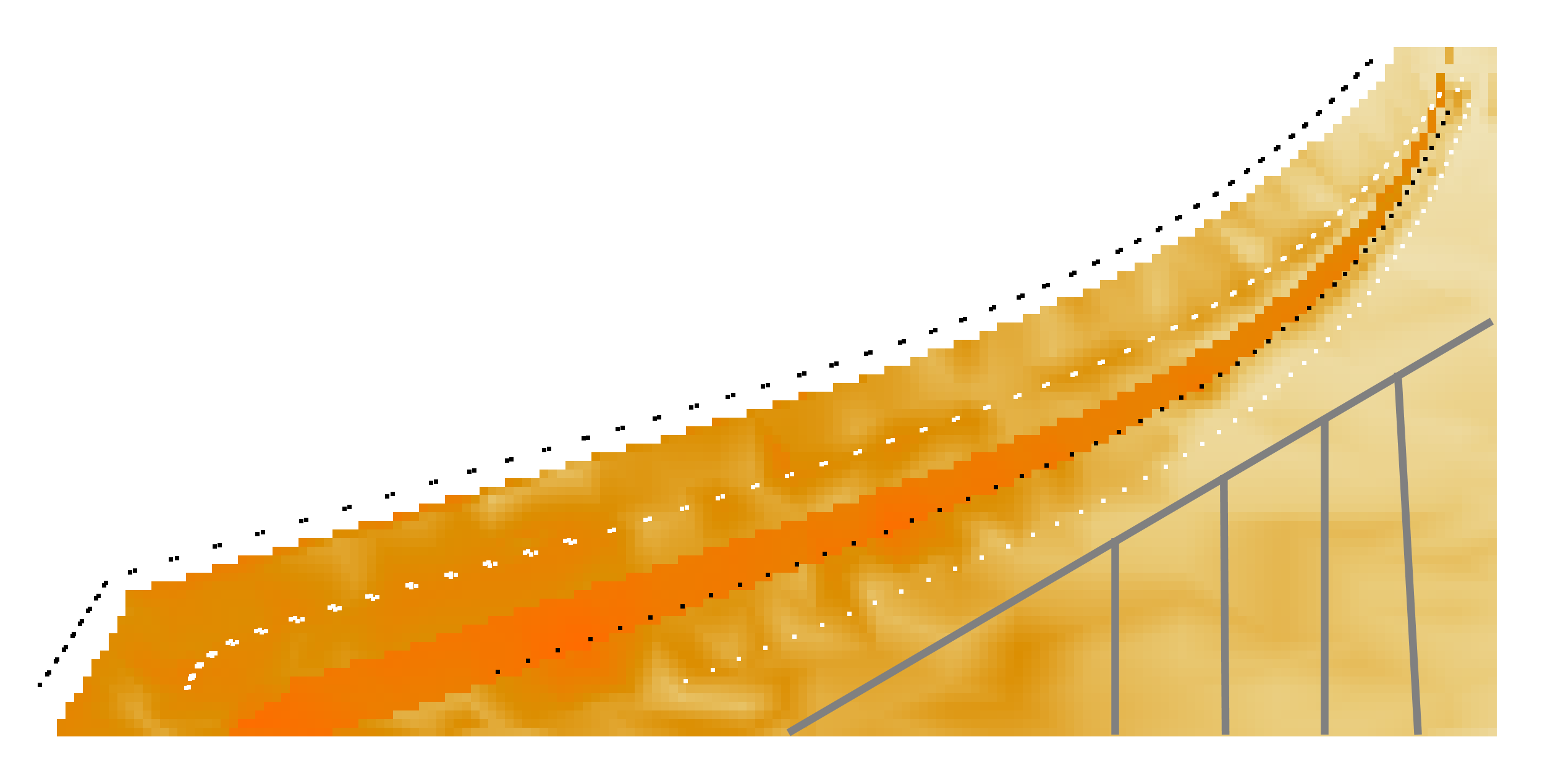}
	\caption[Dulcken:ModelInput]{Thickness (orange background), layout of the 8' and 4' bridges pins (white), 8' and 4' hitchpins (black), and ribs and cutoff bar (gray) of the Dulcken soundboard as input to the FDTD physical model.}
	\label{fig:dulckensoundboarddiscrete}
\end{figure}

The input to the FDTD model is shown in Fig. \ref{fig:dulckensoundboarddiscrete}, displaying the layout of the discretized Dulcken soundboard, consisting of 72 $\times$ 178 nodal points on a regular grid with nodal distances of 1 cm in x- and y-direction. The thickness distribution d(x,y) (orange) was measured at points on the instrument and interpolated using a B-Spline interpolation algorithm  to d(x,y) positions. The 4' hitchpin rail of a thickness of 1 cm was added according to a CAD file provided by the museum, which again was created from a x-ray picture of the instrument.

Typical material parameters for spruce were assumed to be Young's modulus in grain direction (along the strings) at $E_x$ = 11 GPa, a ratio of Young's modulus to the cross-grain direction of $E_x / E_y$ = 8 was assumed. The density was taken as $\varrho = 430 kg/m^3$, and a Poisson ratio of 0.3 was used. These values were also used for the bridges and bars.

The damping is assumed to be a simple constant, which is a first-order differential of the displacement with respect to time multiplied by a constant. In terms of FDTD implementation this results in a logarithmic decrement in the iteration. The increment needs to take the sampling rate of simulation into consideration. Such a damping results in an exponential temporal decay of a knocking sound used as impulse input. It also leads to a stronger exponential damping of higher frequencies compared to lower ones. This damping is way too simple to allow for a realistic modeling of the instrument, as complex viscoelastic effects are omitted. A realistic model would adjust the length of the impulse response using this logarithmic decrement and then damp higher frequencies according to the more complex viscoelastic properties of the wood species. This is beyond the scope of the present paper and could only be performed if the frequency-dependent change of internal damping of wood through aging would already be known, which is not the case. Still, as discussed above, the change of internal damping of the lowest mode was extensively measured in the literature and therefore can be modelled using the logarithmic decrement. The resulting impulse response sound, lacking of viscoelastic damping of higher frequencies, is therefore too bright, still the general trend of damping can be estimated varying the logarithmic decrement, or damping constant.

\subsection{Internal damping analysis}

The internal damping was varied around the measured value, a damping too low $I_{lower}$ = 0.9999 and damping too high $I_{higher}$ = 0.9998. With both steps impulse responses were simulated using the FDTD model at all 8' and 4' string positions on the bridges, resulting in 104 impulse responses per damping case.

From these impulse responses two parameters were derived. Firstly, the decay time T60 was calculated, the time the impulse response needs to decay from 0 dB to -60 dB. This is a common measure in room acoustics, where concert halls or rooms in general are characterized through their reverberation time T60. This measure allows us to compare the modeling results to measured impulse responses on the instrument which have again been performed on all 8' and 4' string positions at the real 1755 Dulcken instrument. These measured T60 allow the estimation of the internal damping the instrument has today. We can then assume that the internal damping might have varied over time and estimate its sound by relating the modeling results for the measured T60 with modeling results for altered T60. The T60 measured with the piano includes internal as well as external damping caused by radiation. Still, with wood, external energy loss is low compared to internal viscoelastic damping making this estimation reasonable.

To estimate how the perceptual difference between today's and original sounds might have been, all modelled impulse responses were calculated in term of their spectral centroid, therefore their brightness. This is the main perceptual parameter immediately perceived by listeners\cite{Bader2013}. We can therefore reconstruct the sound of the 1755 Dulcken from its today's sound in terms of its historical behaviour, at least for this most prominent perceptual feature. 

\vspace{2cm}

\section{Results}

\subsection{Internal Damping}

The measurements of the impulse response T60 have been calculated from the 8' bridge, with three impulses performed on each string position to have a mean of T60 = 235 ms with a standard deviation of 6 ms. If the soundboard had decreased damping through aging we would expect to historically have a shorter T60 when the instrument was built in 1755.

The FDTD simulation of the impulse responses was therefore performed for $I_{lower}$ and $I_{higher}$, a lower and a higher damping compared to the measured case. The values used resulted in T60$_{lower}$ = 306 ms and T60$_{higher}$ = 163 ms, roughly 30\% lower and higher.

\begin{figure}
	\centering
	\includegraphics[width=1\linewidth]{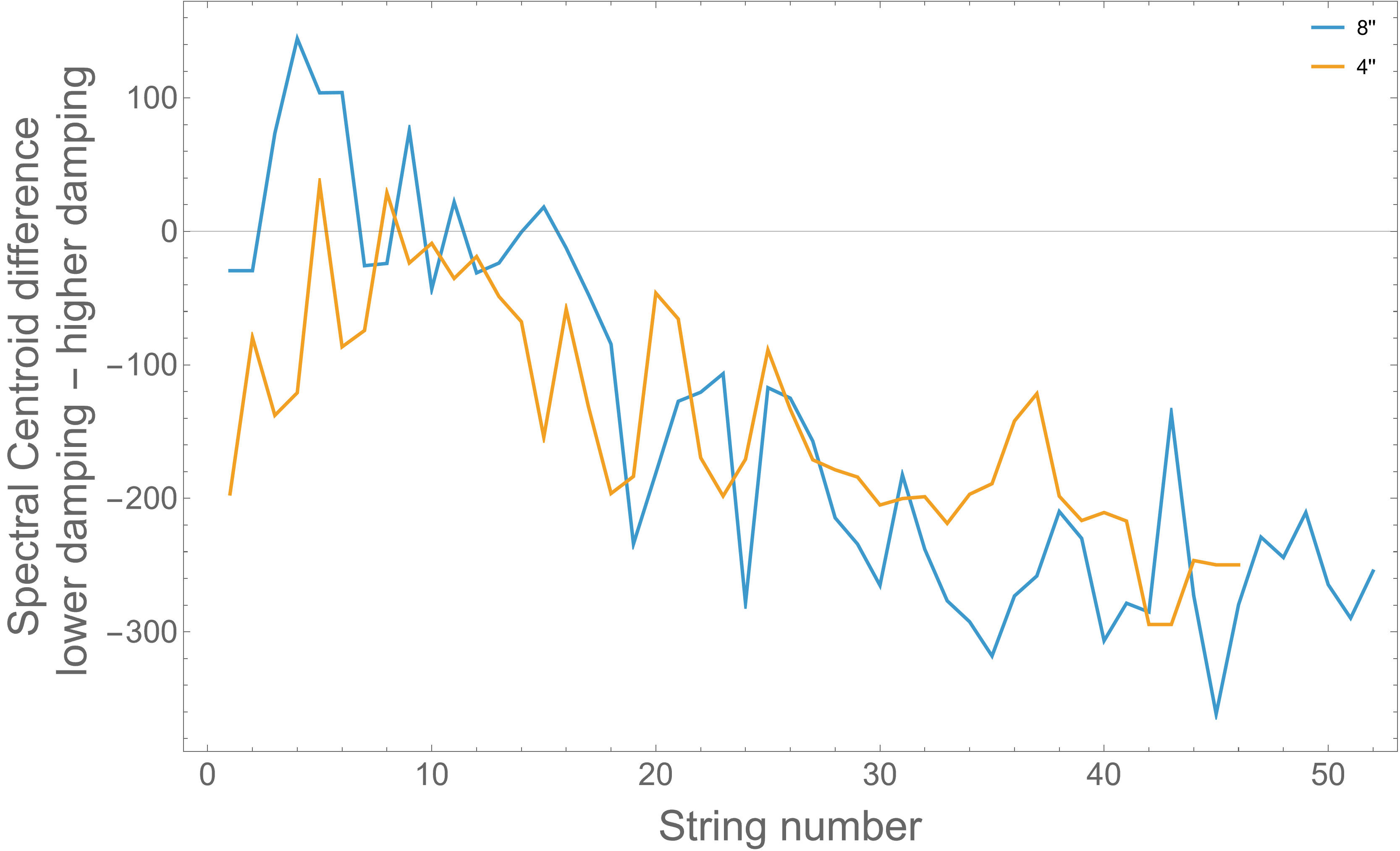}
	\caption{Difference in spectral centroid between the case of increased and decreased damped soundboard through aging of the 1755 Dulcken harpsichord FDTD model for the 8' (blue) and the 4' (yellow) string positions from bass to treble. When the damping had decreased through ages, the brightness of the sound decreases, too. The highest values of the 4' hitchpin rail are simulation artifacts and therefore omitted here.}
	\label{fig:dulckendamping}
\end{figure}

To estimate if an in- or decreased internal damping of a sound is perceived with varying internal damping and string position, the spectral centroids were calculated as a psychoacoustic measure of brightness of the simulated impulse responses. Their differences in terms of changed internal damping are shown in Fig. \ref{fig:dulckendamping} for the 8' (blue) and the 4' (yellow) bridge for all string positions of the 8' bridge and up to position 99 for the 4' bridge, as the remaining string positions had diverged in the calculation.

The overall trend is clearly observable: with decreased damping the instrument sounds less bright. This counterintuitive behaviour can be explained through the decreased filter effect of increased damping discussed above, which mainly appears with high frequencies and have not been taken into consideration only until recently.

It is also of interest that the lower brightness strongly depends on the played keys. While in the low register of the 8' bridge a decreased internal damping indeed corresponds to a increased brightness, as can be found in older studies only taking this region into account, for higher keys the brightness decreases with decreased internal damping. This holds both for the 8' and the 4' bridge pointing to a behaviour not corresponding to absolute frequency but to the position of the strong on the soundboard. Around key 10 nearly no damping change is present where both effects counterbalance each other.

Therefore we can conclude that the effect of changes in damping due to aging alters the brightness of harpsichords such that higher keys become less bright, lower keys become brighter and a middle region is nearly uneffected.

\subsection{Helmholtz resonance}

The Helmholtz resonance frequency was extracted from the impulse response measurements to be at 37 Hz. The calculated value, taking the soundhole radius and air volume into consideration is also 37 Hz, so in very good agreement with the measurement.

\subsection{Soundboard stress reduction due to double bentside}

The Dulcken harpsichord has the 8' strings attached to the outer wall of a double bentside construction. The soundboard is glued to the inner wall. On the curved part of the wall and the small back part of the two walls a gap of air avoids contact between the two walls. Nevertheless, the x-ray images show nails connecting both walls. The long walls are connected one to another. Also both walls are glued to the bottom plate.

Three reasons for this construction are mentioned. It is assumed that attaching the 8' strings to the outer wall, the static stress of the strings on the soundboard is eased. Furthermore, this construction is known to arrive at a higher tuning stability. Lastly, the sound was described as more bright compared to a attaching the strings to the inner wall as usual with harpsichords.

\begin{figure}
	\centering
	\includegraphics[width=.7\linewidth]{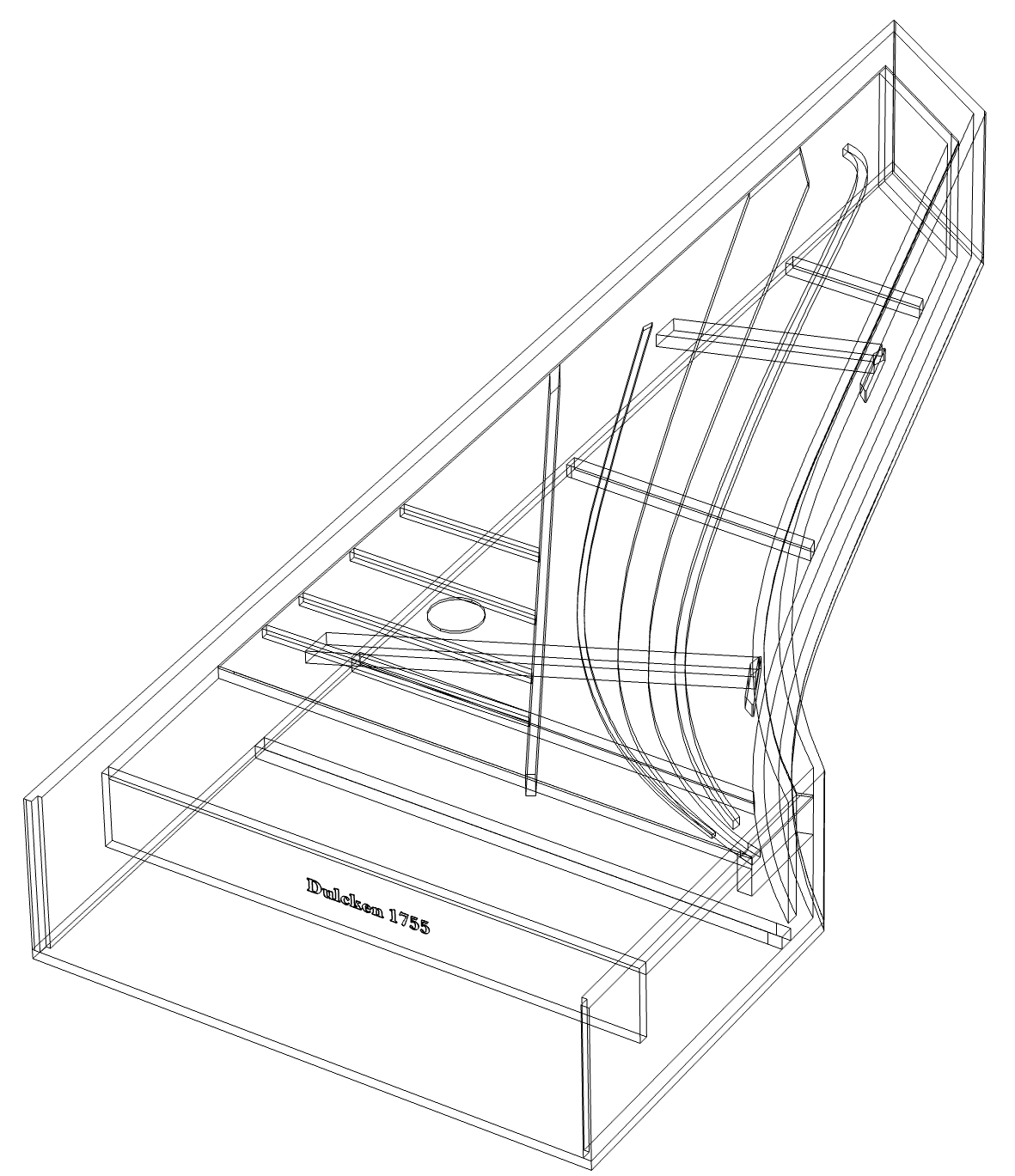}
	\caption{Geometry of the CAD model of the Dulcken harpsichord used for Finite-Element Method (FEM) stress analysis.}
	\label{fig:dulckengeometrie}
\end{figure}

To arrive at an estimation of stress distribution on a harpsichord soundboard, a 3D CAD model of the Dulcken has been constructed from the several sources mentioned above as shown in Fig. \ref{fig:dulckengeometrie} . This model includes, next to the soundboard, the two walls, the bottom the front plate as well as the bars in the instrument enhancing stability. For the entire instrument, material parameters of spruce was taken as wood for all parts. The soundboard thickness was simplified to 3 mm thickness throughout to enhance calculation stability.

The forces of the strings acting on the instrument were calculated from the present stringing and tuning at a' = 392 Hz. The 8' strings are brass from the lowest key at HH to B and ion from H to d'''. The 4' strings are brass from H to f and ion from there on. The lowest 8' string has a diameter of 0.5 mm while the highest has 0.2 mm. The lowest 4' string diameter is 0.3 mm while the highest is 0.2 mm again. Each of the 52 keys are double stringed in 8' and single stringed in 4' resulting in 156 strings altogether. The strings scaling were taken from measurements at the instrument. Taking typical densities for brass of $\varrho = 8.635 g/cm^3$ and iron of $\varrho = 7.874 g/cm^3$ the force for each string needed to arrive at the respective pitch was calculated. 

\begin{figure}
	\centering
	\includegraphics[width=1\linewidth]{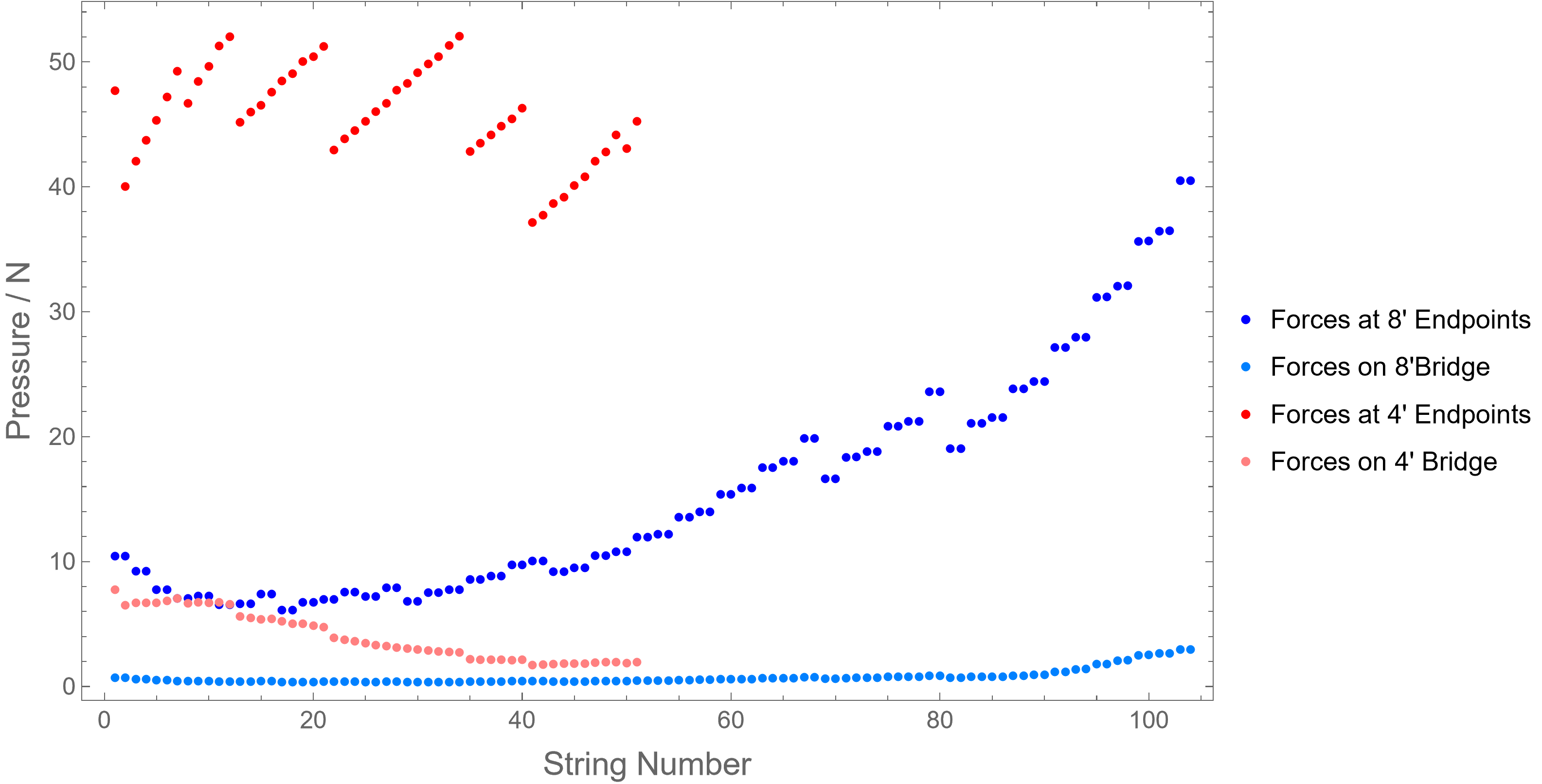}
	\caption{String forces for the 8' (blue) and 4' (red) strings. Dark color: Forces necessary to arrive at the respective pitches the instrument is tuned in its present stage using string material (brass, iron), string thickness, and measures (diapason). Light color: Forces acting on the bridges. The highest forces are found with the 4' strings which are acting on the 4' hitchpin rail on the soundboard. The total force acting on the instrument is 3938 N, roughly corresponding to 401 kg.}
	\label{fig:stringtension}
\end{figure}

Fig. \ref{fig:stringtension} shows the 8' (blue) and 4' (red) forces of the strings acting on the x-direction at the hitch pins (dark blue, dark red) and on the bridges (light blue, light red). The bridge forces also act on the hitch pins but in the z-direction. With the stringing used at present, by far the most forces are applied to the 4' hitch pin rail on the soundboard. Only the high 8' strings add such strong forces but do so to the outer wall. The forces acting on the bridges are much smaller, as expected due to the small angle the strings have over the bridge. Of course, the downward force on the bridge is only an increment of the tension it is tuned with. In total 3938 N are acting on the instrument, corresponding to about 401 kg.

Using the angles of the strings over the 8' and 4' bridges, the forces normal or perpendicular to the soundboard were derived. The forces at the hitch pins are attached to the respective geometries as forces acting in the long or x-direction in-plane to the soundboard and forces acting in the normal or z-direction of the soundboard. Including these forces in the FEM model has the advantage that each of them can be turned on and off at need to arrive at estimations of the amount of stress added to the soundboard due to one of these string force parts.

To arrive at an estimation of the impact of the double bentside, the 8' string ends were attached to the outer and to the inner wall such that the model can easily switch between both versions.

\begin{figure}
	\centering
	\includegraphics[width=1\linewidth]{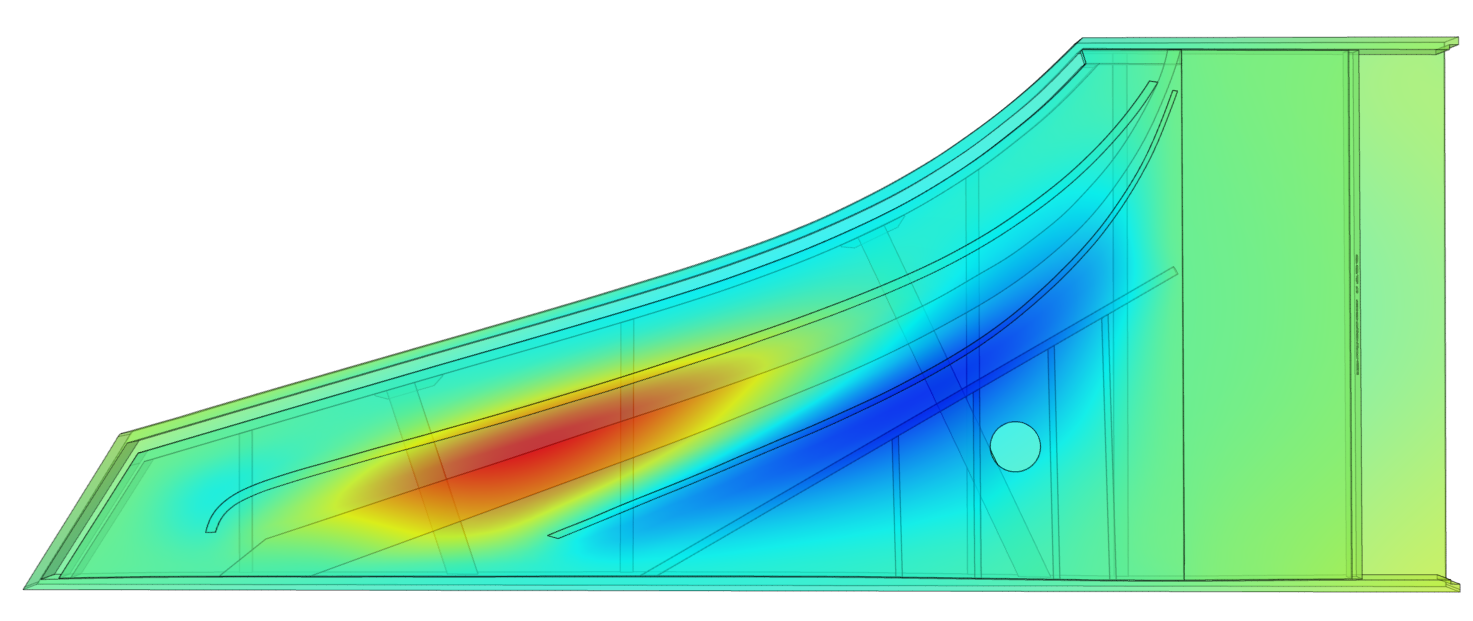}
	\caption{Static displacement of the Dulcken soundboard due to string forces. Red is outward, blue is inward buckling. The maximum displacement is $\pm$ 0.15 mm}
	\label{fig:dulckenallaussendisp}
\end{figure}

Fig. \ref{fig:dulckenallaussendisp} shows the displacement of the instrument due to the forces of the strings acting on the different parts. Here, the 8' strings are attached to the outer wall. The buckling is outwards (red) around the hitch pin rail and inward around the 4' bridge as expected. The maximum displacement is $\pm$ 0.15 mm which is very low compared to the soundboard size. Comparing the maximum buckling to the case where the simulation was run with switching off the 4' hitch pin forces a maximum displacement of 7 mm was found! Therefore, the present construction very well keeps the balance which is not astonishing as the same force acting on the hitch pins in the upward z-direction is acting on the 4' bridge downwards. The result is the well-known buckling 'wave' where very old soundboards tend to flow into in a plastic manner.

\begin{figure}
	\centering
	\includegraphics[width=1\linewidth]{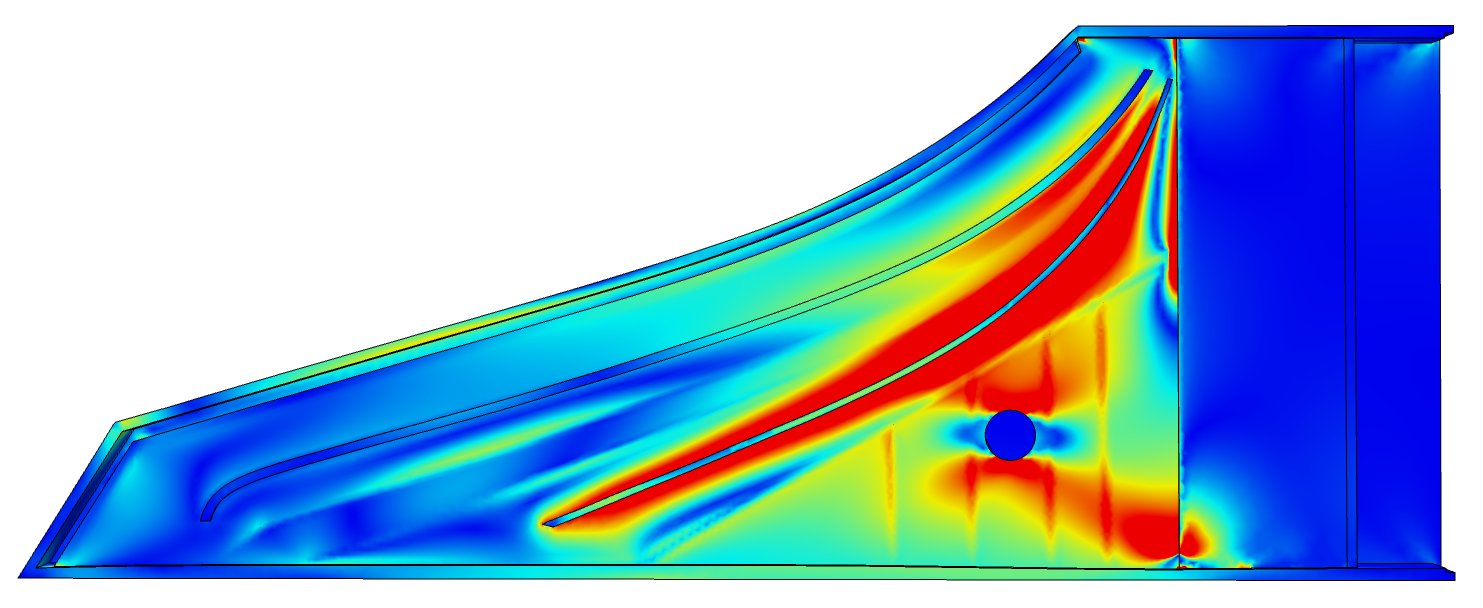}
	\includegraphics[width=1\linewidth]{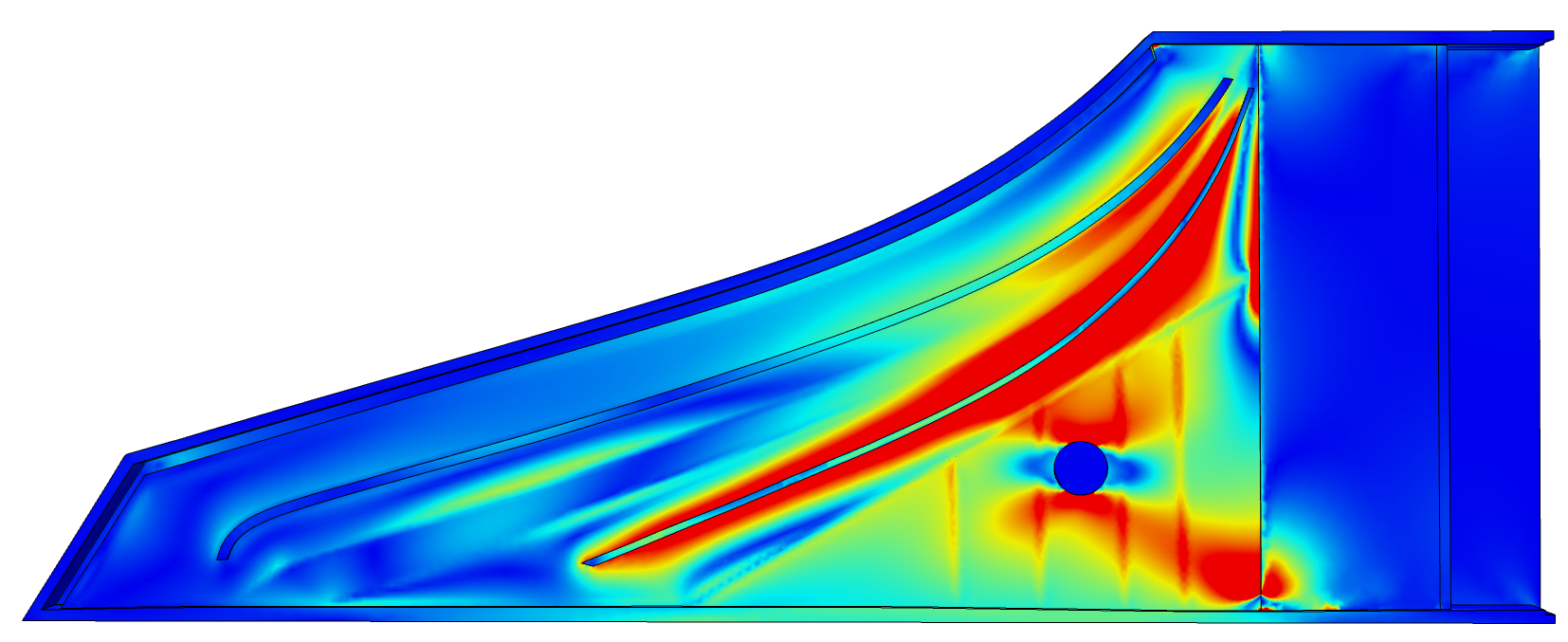}
	\caption{Comparison of the tension on the soundboard with 8' strings attached to the outer (upper plot) and inner (lower plot) wall.}
	\label{fig:dulckenallaussen}
\end{figure}

Of much more interest, compared to the buckling, is the stress, the tension in the soundboard when the 8' strings are attached to the outer or inner wall. Fig. \ref{fig:dulckenallaussen} compares the two cases where nearly no difference can be seen. This is confirmed in Tab. \ref{tab:tensionscases} which shows the tensions integrated over the whole soundboard for several cases. All tensions are calculated as $\sqrt{s_x^2+s_y^2+s_z^2}$ with the tensions $s_x, s_y,$ and $s_z$ in the x-, y-, and z-direction respectively. In the first two rows tensions are shown when all strings act on all parts, so the real instrument, in the case of the 8' strings attached to the outer wall with a stress of 1530 $N/m^2$ and when the 8' strings are attached to the inner wall of 1458 $N/m^2$. The tensions are practically the same.

The reason for this astonishing finding is that the tension of the outer wall is acting on the bottom plate of the instrument. From there it is transferred to the inner walls and from there to the soundboard.  

\begin{figure}
	\centering
	\includegraphics[width=1\linewidth]{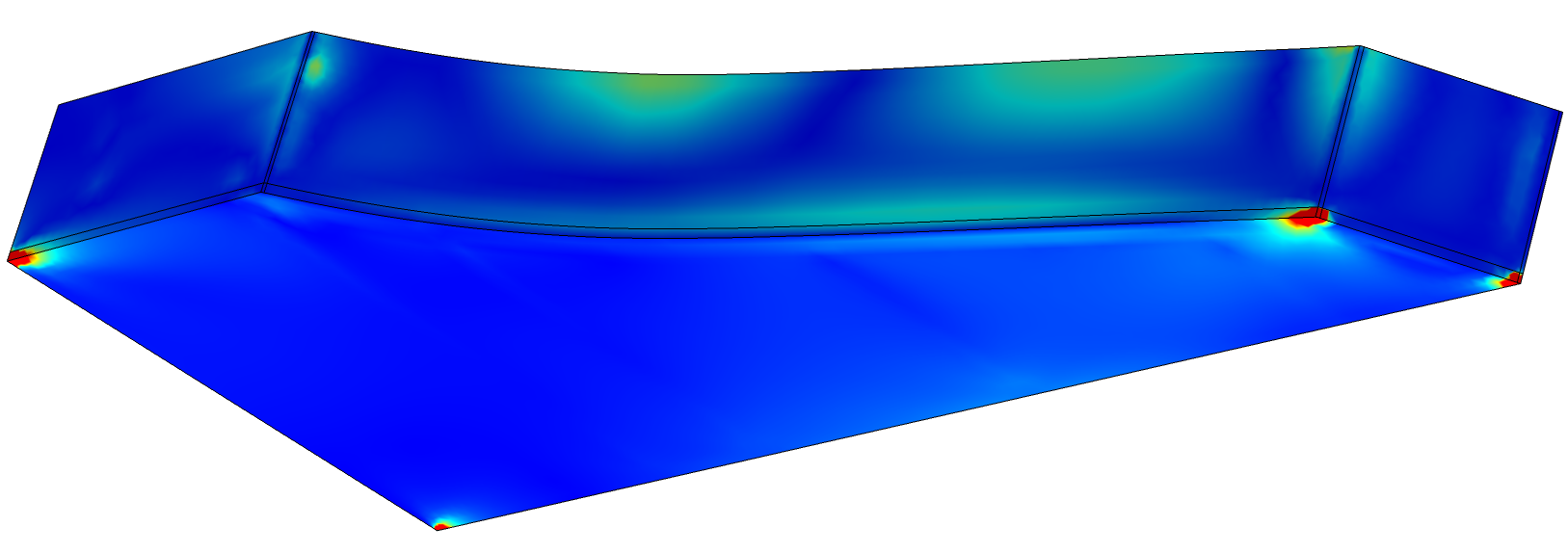}
	\includegraphics[width=1\linewidth]{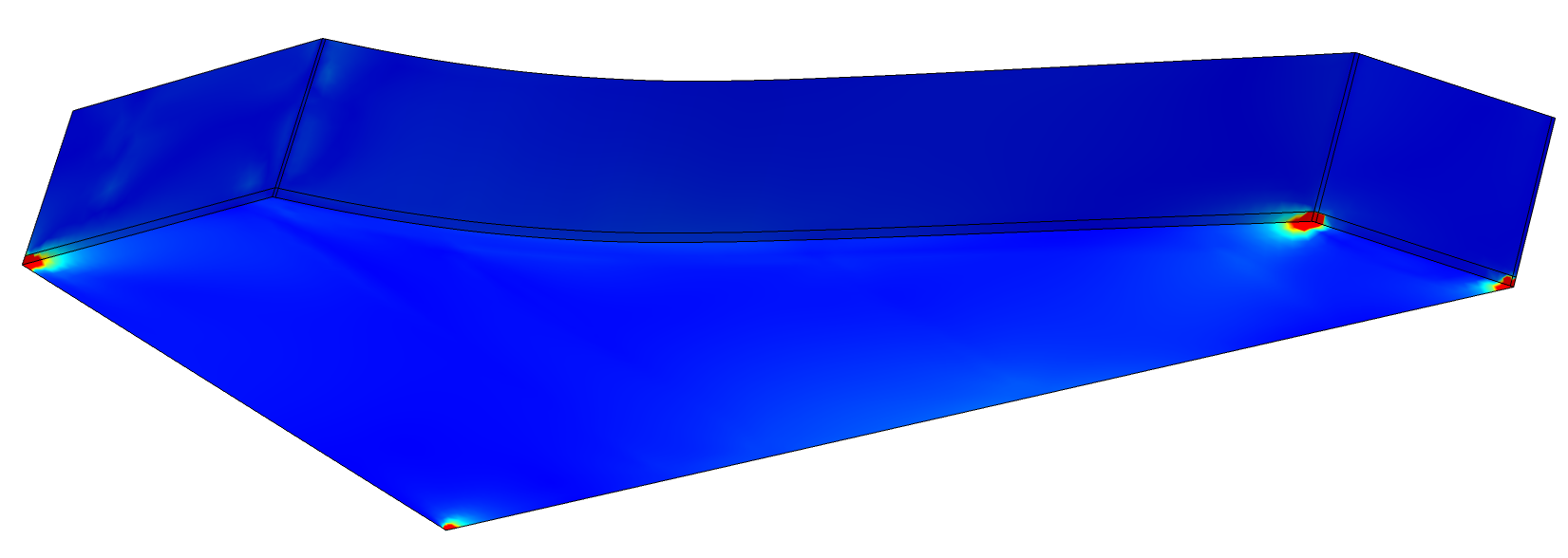}
	\caption{Stress distributions on the back of the harpsichord comparing the cases of 8' strings attached to the outer (top plot) or inner (bottom plot) wall. In the outer wall case the tension is transferred to the soundboard via the bottom plate of the instrument.}
	\label{fig:dulckennur8aussen}
\end{figure}

Tab. \ref{tab:tensionscases} furthermore displays the integrated stress of the whole soundboard for different applied forces on the instrument turned on or off. The strongest stress is applied by the 4' hitch pins with 1189 $N/m^2$. As this force is acting mainly in the in-plane direction, the soundboard is much more able to withstand such forces compared to forces acting normal to it. Next is the 4' bridge caused by the high tension of these strings. Only then follows the outer and inner walls, about the same between 640 and 550 $N/m^2$. It is astonishing that the outer wall 8' attachment leads to a bit more stress than the inner wall attachment.

\begin{table}
	\centering
\begin{tabular}{|c|c|}
	\hline
Body	& stress [$N/m^2$] \\
	\hline
All forces, 8' outer wall & 1530 \\ 
	\hline
All forces, 8' inner walll & 1458	\\
	\hline
Only 8' outer wall & 640 \\
	\hline
Only 8' inner wall &  550 \\
	\hline
Only 8' bridge	& 222 \\
	\hline
Only 4' bridge	&  934 \\
	\hline
Only 4' rail	& 1189 \\
	\hline
Only 8' \& 4' bridge and 4' rail	& 1099 \\
\hline
\end{tabular}
\caption{Stress integrated over soundboard for 8' strings attached to the outer or inner wall. No significant difference on soundboard tension appears. Other combinations show strongest soundboard stress caused by 4' hitch pin rail.} 
\label{tab:tensionscases}
\end{table}

Both cases of the 8' strings attached to the outer and inner wall are compared in Tab. \ref{tab:tensionsparts}. The whole tension splits into components of the plain soundboard, the bridges, the 4' hitch pin, cutoff bar, and ribs. Of course the plain soundboard is the largest geometry and therefore takes most of the stress. It is interesting to see that the ribs behind the cross bar only contribute a small amount. The rail and cross bar stabilizes the soundboard considerably, followed by the 8' and 4' bridges.

\begin{table}
	\centering
\begin{tabular}{|c|c|c|c|c|c|c|c|}
	\hline
				& Soundboard  & 8' bridge & 4' bridge  & 4' hitch pin rail  & cross bar  & ribs   & total\\
	\hline
8' outer wall	& 1140		  & 61 	  & 19  	   & 226 			   & 70 		& 14    & 1530\\
				& 74.5\%      & 4.0 \%    & 1.2 \%     & 14.7 \%		   & 4.7 \%	    & 0.9 \% &      \\
	\hline
8' inner wall	& 1088		  & 52 	  & 18 	   & 219 			   & 67 		& 14    & 1458\\
			    & 74.6 \%	  & 3.6 \%    & 1.2 \%     & 15.0 \%		   & 4.6 \%     & 1.0 \% &       \\
			     \hline		   
\end{tabular}
\caption{Stress[$N/m^2$] integrated over different soundboard parts when 8' strings are attached to outer or inner wall. The amount of stress in the parts roughly corresponds to the size of these parts.}
\label{tab:tensionsparts}
\end{table}

\section{Conclusions}

Frequency-dependency of damping is well known. Higher frequencies are damped stronger than lower ones. Also due to the complexity of wood, this frequency-dependent damping is not simply an exponential decay but has a much more complex structure, giving different wood types different characteristics. Yet, a third frequency-dependency of damping is to be considered as shown above, the brightness change due to damping depends on the position of the string on the soundboard. This finding is hard to validate experimentally, as we would need to measure the harpsichord over a considerable aging period. Still the main behaviour needs to be taken into consideration and therefore the simple conclusion that old musical instruments sound more or sound less bright compared to new ones need to be more differentiated.

The construction of the outer wall displaced from the inner wall and the attachment of the 8' strings on the outer wall does not lead to the expected reduction of tension within the soundboard as the tension is transferred to the soundboard via the bottom of the instrument. This finding is not even taking the nails into consideration which attach the outer and inner walls at some locations. Whether this construction leads to other advantages is not part of the present study.

The present stringing of the instrument shows much higher tensions in the 4' strings compared to the 8' strings. This is expected, causing the tension distribution in the instrument to be one dominated by the 4' strings.

Future studies need to go into details of viscoelastic effects which result in frequency band-gaps disturbing the simple exponential decay of the spectrum\cite{Martinez2024}. This topic is also of interest in terms of wood species substitution caused by wood scarcity due to climate change. When replacing wood, achieving a similar eigenspectrum as traditional wood can be achieved by changed the thickness distribution of the soundboard. Still, this will not lead to the same sound as wood types have different internal damping and therefore each species has its characteristic sound.

\newpage
\renewcommand{\notesname}{Endnoten}
\renewcommand{\enotesize}{\normalsize}
\theendnotes


\begin{thebibliography}{99}
	
\bibitem{Bader2022} Bader, R.: Impact of Damping on Oscillation Patterns on the Plain Piano Soundboard. Acoustics 2022, 4, 1013–1027. https://doi.org/10.3390/ acoustics4040062	

\bibitem{Bader2021} Bader, R.: How Music Works – A Physical Culture Theory. Springer 2021.



\bibitem{Bader2013} Bader, R..: Nonlinearities and Synchronization in Musical Acoustics and Music Psychology. Springer Series Current Research in Systematic Musicology, Vol. 2, Springer Heidelberg , 2013.

\bibitem{Bader2005} Bader, R.: \emph{Computational Mechanics of the Classical Guitar,} Springer 2005.

\bibitem{Beurmann2010} Beurmann, A., Bader, R., \& Schneider, A.: An acoustical study of a Kirkman harpsichord from 1766. Galpin Society Journal LXIII, 61-72, 2010.

\bibitem{Fletcher2000} Fletcher, N. \& Rossing, Th.: Physics of Musical Instruments. Springer 2000.

\bibitem{Gurau2023} Gurau, L., Timar, M. C., Cosereanu, C., Cosnita, M., \& Stanciu, M. D.: Aging of Wood for Musical Instruments: Analysis of Changes in Color, Surface Morphology, Chemical, and Physical-Acoustical Properties during UV and Thermal Exposure, Polymers 15, 1794, 1-22, 2023.

\bibitem{Holz1973} Holz, D.: Untersuchungen an Resonanzholz. 5. Mitteilung: Über bedeutsame Eigenschaften nativer Nadel- und Laubhölzer in Hinblick auf mechanische und akustische Parameter von Pianoresonanzböden. [Investigations on resonance wood. 5. Report: On crucial properties of native firious and broad-leaved wood with respect to mechanical and acoustica parameters of piano soundboards.] Holztechnologie 14, 195-202, 1973.

\bibitem{Kou2015} Koe, Y., Liu, B. \& Tian, J.: Radiation efficiency of damped plates. J. Acoust. Soc. Am. 137, 1032-1035 (2015).

\bibitem{Larisch2020} Larisch, L., Lemke, B. \& Wittum, G.: 3d Modeling and Simulation of a Harpsichord. Computing and Visualization in Science 23(1-4), 2020.

\bibitem{LeMoyne2012}Le Moyne, S., Le Conte, S., Olivier, F., Frelat, J., Battault, J.-C., \& Vaiedelich, S.: Restoration of a 17th-century harpsichord to playable condition: A numerical and experimental study. J. Acoust. Soc. Am. 131(1), 888-896, 2012.

\bibitem{Loredo2011} Loredo, A., Plessy, A., Hafidi, A. E. \& Hamzaoui, N.: Numerical vibroacoustic analysis of plates with constrained-layer damping patches. J. Acoust. Soc. Am. 129, 1905-1918 (2011).

\bibitem{Noguchi2008} Noguchi, T., Obataya, E., \& Ando, K.: Effects of aging on the vibrational properties of wood, J. Cultur. Herit., 13S, 21-25, 2012.

\bibitem{Martinez2024} Martinez, C. \& Bader, R.: Machine Learning-Analysis of a Physics-informed Dataset of a Viscoelastic Damped Membrane. DAGA Proceedings 983-986, 2024.

\bibitem{Ono1996} Ono, T.: Frequency responses of wood for musical instruments in relation to the vibrational properties. J. Acoust. Soc. Jpn. (E) 17,4, 183-193, 1969.

\bibitem{Obataya2020} Obataya, E., Zeniya, N., \& Endo-Ujiie, K.: Effects of seasoning on the vibrational properties of wood for the soundboard of string instruments. J. Acoust. Soc. Am. 147 (2), 998-1005 , 2020.

\bibitem{Obataya2017}  Obayata, E.: Effects of natural and artificial aging on the physical and acoustic properties of wood in musical instruments, J. of Cultural Heritage 27S, 63-69, 2017.

\bibitem{Penttinen2006} Penttinen, H.: On the dynamics of the harpsichord and its synthesis. Proceedings DAFx-06, 2006.

\bibitem{Plath2017} N. Plath, F. Pfeifle, Ch. Koehn, and R. Bader: Radiation
Characteristics of Grand Piano Soundboards in Different Stages of Production,
Proceedings of the 2017 International Symposium on Musical Acoustics, Montreal, Canada, 18-22, 2017.

\bibitem{Plath2019} Plath, N.: From workshop to concert hall: acoustic observations on a grand Piano under construction. Diss. Univ. of Hamburg 2019.

\bibitem{Stoel2008} Stoel, B. C. \& Borman, T. M.: A Comparison of Wod Density between Classical Cremonese and Modern Violins. PLoS ONE 3 (7), 1-7, 2008.Und

\bibitem{Underwood1992} Underwood, J. H., Burr, W., Kapusta, A. A., \& Rickart, Ch. A.: Characterization of Early and Modern Wire for an Italian
Harpsichord. Am. Soc. for Testing and Materials, 20(4), 312-317, 1992.

\bibitem{Välimäki2004} V\"alim\"aki, V., Penttinen, H., Knif, J., Laurson, M., \& Erkut, C.: Sound Synthesis of the Harpsichord Using a Computationally Efficient Physical Model, EURASIP Journal on Advances in Signal Processing 2004, 2004. 

\bibitem{Waltham2008} Walthman, Ch. \& Kotliki, A.: Vibrational characteristics of harp soundboards. J. Acoust. Soc. Am. 24(3), 1774-1780, 2008.

\bibitem{Yokoyama2009} Yokoyama, M., Gril, J., Matsuo, M., Yano, H., Sugiyama, J., Clair, B., Kubodera, S., Mistutani, T., Sakamoto, M., Ozaki, H., Imamura, M., \& Kawai, S.: Mechanical characteristics of aged Hinoki wood from Japanese historical buildings, C. R. Physique 10, 601-611, 2009.

\end{thebibliography}
  \end{document}